\documentclass[aps,prd,onecolumn,nofootinbib,superscriptaddress,showpacs,amsmath,amssymb]{revtex4}
\usepackage{graphicx,epsf}
\usepackage[]{latexsym}

\newcommand{\be}{\begin{eqnarray}}
\newcommand{\ee}{\end{eqnarray}}
\newcommand{\bea}{\begin{eqnarray}}
\newcommand{\eea}{\end{eqnarray}}

\def\comment#1{}

\newcommand{\mpl}{m_{P}}

\usepackage{color}

\definecolor{darkred}{rgb}{.8,0,0}

\definecolor{darkblue}{rgb}{0,0,.7}

\definecolor{darkgreen}{rgb}{0,.7,0}

\begin{document}

%
%
\title{Uncertainty principle and gravity\footnote{presented at the Conference ``Eternity between Space and Time'',  Padova, May 19-21, 2022.}}

\author{Fabio Scardigli}
\email{fabio@phys.ntu.edu.tw}
\affiliation{Dipartimento di Matematica, Politecnico di Milano, Piazza Leonardo da Vinci 32, 20133 Milano, Italy}
\affiliation{Department of Applied Mathematics, University of Waterloo, Ontario N2L 3G1, Canada}
%
%
%
%

\begin{abstract}
\par\noindent
We give a pedagogical introduction to the generalized uncertainty principle (GUP), by showing how it naturally emerges when the action of gravity is taken into account in measurement processes. We review some physical predictions of the GUP. In particular we focus on the bounds that present experimental tests can put on the value of the deformation parameter $\beta$, and on the prediction of black hole remnants, which are a good candidate for dark matter. In passing, we also quote a theoretical value computed for $\beta$, and swiftly comment on the vast parameter region still unexplored, and to be probed by future experiments.   

\end{abstract}
%

%
\maketitle
\section{Introduction}
\noindent Uncertainty principle is the cornerstone of Quantum Mechanics~\cite{Heisenberg27}. The principle, when referred to single point particles, can be formulated in a quite simple way:\\

\noindent \textit{It is impossible to measure, with an arbitrary precision, position and speed of a particle at the same instant in time.}\\   

\noindent The idea emerged when Heisenberg, in a seminal paper of 1927, discussed measurement processes in Quantum Theory. He illustrated the principle through a gedanken experiment, which from then is known under the name of `Heisenberg microscope argument'~\cite{Heisenberg30}. Heisenberg's original idea was to measure position and momentum of a static particle, say an electron, by using a photon as a probe. The photon scatters off the electron, and by measuring the properties of the photon after the scattering, one would like to know the exact position $x_e$ and momentum $p_e$ of the electron at the instant of the scattering. However, since the photon has a wavelength 
$\lambda$, from the principles of wave optics follows that the uncertainty in the position of the electron is (at least) $\Delta x_e \simeq \lambda$. Moreover, the photon carries a momentum $p=E/c=h/\lambda$, which, during the scattering, is partially transferred to the electron in an unknown magnitude and direction. This implies that, just after the scattering, the uncertainty in the electron momentum amounts to (at most) 
$\Delta p_e \simeq p = h/\lambda$. Therefore, Heisenberg concluded that
\be
\Delta x_e \, \Delta p_e \simeq \lambda \cdot \frac{h}{\lambda} \simeq h
\ee
Successively, Robertson and Schr\"odinger~\cite{Robertson1929,Schr1930} formulated the uncertainty principle for canonically conjugated variables, such as the position $x$ and momentum $p$ of a particle, in the form
\be
\Delta x \, \Delta p \geq \frac{\hbar}{2}\,,
\ee
which is the expression commonly accepted today.\\
In those early approaches the gravitational interaction between particles was completely neglected, although this attitude was somehow justified by the huge weakness of gravity, when compared with other fundamental interactions. However, Heisenberg's heuristic approach paved the way to the formulation of a generalized uncertainty principle, 
GUP,~\cite{Bronstein36,Amati87,Maggiore93,Kempf95,Scardigli99,Adler99,Capozziello00,Scardigli03}, which originate just by properly taking into account the gravitational effects in the photon-particle interaction. For example, the unavoidable formation of (microscopic) black holes in the measurement process, or even the simple Newtonian photon-electron gravitational attraction, imply the existence of a \textit{minimum testable length}, below which position measurements become meaningless.


\section{Generalized uncertainty principle} 
\label{GUP}
\noindent

A possible generalization of the classical Heisenberg argument goes roughly as follows~\cite{Scardigli99}. According to Heisenberg principle, one might probe a length as small as one wishes, provided one uses enough energetic photons. In fact, the size of the observable detail is the same as the wavelength of the photon, 
$\delta x \simeq \lambda$. But a photon of wavelength $\lambda$ brings an energy $E=hc/\lambda$, therefore
\be
\delta x \simeq \frac{hc}{E}\,.
\label{deltaxH} 
\ee
So that, increasingly large energies allow to explore decreasingly small details. Incidentally, this is why we build particle accelerators ever larger and more powerful. If you want to probe small details, you need a lot of energy! Now, the interaction between a photon and, say, an electron, is punctual, as far as we know, namely the energy $E$ carried by the photon is packed in a region of size at most $\delta x$, at the instant of the interaction. Here gravity enters the game. In general, we know that any given mass $M$ has an associated gravitational radius 
\be
R_g = \frac{2GM}{c^2}
\label{Rg}
\ee 
or, in terms of energy, $R_g = 2GE/c^4$. The physical meaning of these equations is that, when a mass $M$, or an energy $E$, is packed inside its own gravitational radius $R_g(E)$, then a black hole is formed. From the sphere of radius $R_g(E)$, the so called event horizon, not even light can escape, so that region appears black. Of course, inside a black hole one cannot watch, by definition! So, every detail which falls inside the sphere of radius 
$R_g(E)$ is invisible for the outside observer. In terms of our measurement process, this means that, if the probing photon is too energetic, a micro black hole of size $R_g(E)$ is formed, and this results into an unobservable region of size $R_g(E)$. Essentially the event horizon screens smaller details in this region, and this results into a further error that should be added to the usual quantum mechanical error predicted by the standard Heisenberg principle. Once the micro black hole is formed, even if we try to increase the photon energy $E$ to "see" more refined details (according to the equation $\delta x \simeq hc/E$), what we are really doing is just "to feed" the micro black hole. It will grow, and will hide more and more details behind its event horizon of radius $R_g(E)$. The above considerations can be summarized in a mathematical formula. Precisely, the smallest detail that a photon of energy 
$E$ can actually distinguish is
\be 
\delta x \ \simeq \ \frac{hc}{E} \ + \ \beta R_g(E)\,.
\label{deltax}
\ee
Here $\beta$ is a dimensionless parameter, which is not in principle fixed by the theory, but it is generally assumed to be of the order of unity. An explicit analytic calculation of $\beta$ in~\cite{SLV} has confirmed the circumstance. However, many studies have appeared in literature, with the aim to set experimental bounds on $\beta$ 
(see, for instance, Ref.~\cite{SC}). For a standard Schwarzschild black hole the gravitational radius is given by Eq.\eqref{Rg}, so Eq.\eqref{deltax} now reads
\be
\delta x \ \simeq \ \frac{hc}{E} \ + \ \beta \,\frac{2GE}{c^4}\,.
\label{deltaxE}
\ee
Since for photons the dispersion relation $E=pc$ holds, and repeating the considerations seen in Sec.1, we can say that the uncertainty in the electron's momentum just after the scattering is, at most, $\Delta p \simeq p = E/c$. So, introducing the definitions of Planck length, and Planck mass, respectively, as
\be
\ell_p^2 = \frac{G\hbar}{c^3}\,, \quad \quad m_p c\,\ell_p = \frac{\hbar}{2}
\ee
we can write the new \textit{Generalized Uncertainty Principle} (GUP) as
\be
\Delta x \Delta p \ \geq \ \frac{\hbar}{2} \left[1 + \beta \left(\frac{\Delta p}{m_p c}\right)^2\right]\,.
\label{gup}
\ee
It is remarkable that many different approaches lead to the same formulation of the GUP. Not only gedanken experiments involving micro black holes, but also arguments considering large black holes~\cite{Maggiore93}, or arguments from String Theory~\cite{Amati87}, and even considerations involving just the simple Newtonian gravity~\cite{Adler99}, they all lead to the same Eq.\eqref{gup}. The diagram in Fig.1 summarizes the situation described by Eqs.\eqref{deltaxE}, \eqref{gup}.
\begin{figure}[t]
	\includegraphics[scale=0.8]{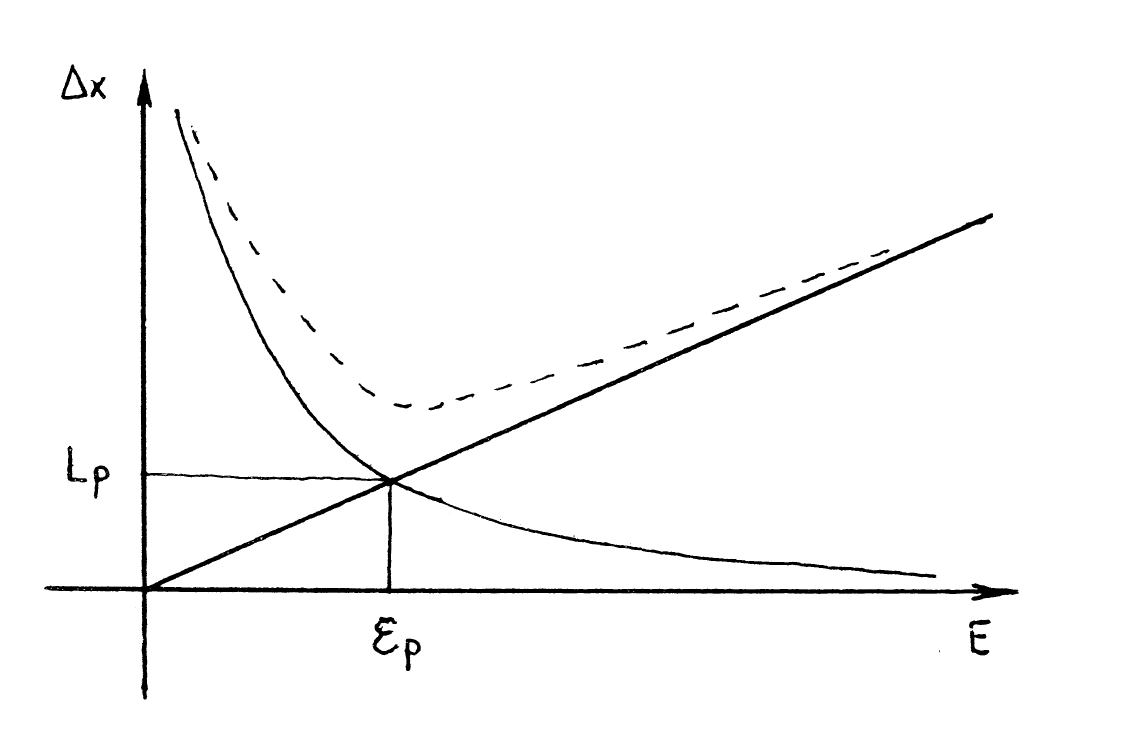}
	\centering
	\protect\caption{Illustrations of Eqs.\eqref{deltaxE}, \eqref{gup}, see text.}
	\label{fig1}
\end{figure}
If gravity did not exist at all, then one could probe lengths small at will, provided to use enough large energies. However, the very existence of gravity, and of a gravitational radius linearly proportional to the energy of the probe itself, modifies the Heisenberg hyperbole with a straight line, and the actual uncertainty $\Delta x(E)$ is described by the dashed line in Fig.1. Thus, Eqs.\eqref{deltaxE}, \eqref{gup} imply the existence of a \textit{minimum observable length} (or detail), which can be easily computed to be $L_{min} \simeq \ell_p\sqrt{\beta}$. Such small details can be observed, in principle, only by using probes (e.g. photons) of Planckian energies, namely 
$E\simeq E_p/\sqrt{\beta} = m_p c^2/\sqrt{\beta}$. To have an idea of how far still we are from reaching those energies, we just remind that the most powerful particle accelerator in the world, the LHC of CERN, in Geneva, reaches at its best $10^4$ GeV, while the Planck energy is at $E_p \simeq 10^{19}$ GeV.

\section{Deformed Quantum Mechanics: tests of GUP}

The generalized uncertainty relation \eqref{gup} can be described also in terms of a fundamental commutator between the position $\hat{X}$ and momentum $\hat{P}$ operators, as
\be
[\hat{X}, \hat{P}] = i\hbar \left(1+ \beta \,\frac{\hat{P}^2}{m_p^2 \,c^2}\right)\,,
\label{gupcomm}
\ee
where the fundamental variables $\hat{X}$, $\hat{P}$ are thought to be the physical high energy operators useful, in particular, to describe physics at or near the Planck scale. In general they can have non-linear representations,
$\hat{X}(\hat{x})$, $\hat{P}(\hat{p})$, in terms of the operators $\hat{x}$, $\hat{p}$ which are the usual position and momentum operators at low energy, obeying the standard Heisenberg canonical commutator 
$[\hat{x},\hat{p}]=i\hbar$. Several authors (see e.g. Refs.~\cite{Kempf95,Brau99,Das08,Bosso22}) have used the above approach to construct a \textit{Deformed (or Generalized) Quantum Mechanics}. In this way it becomes possible to re-compute virtually any quantum phenomenon by using the GUP commutator \eqref{gupcomm}. The results (each one dependent, obviously, on the deformation parameter $\beta$) have been compared with the experimental data, so to extract (upper) bounds on the parameter $\beta$. Besides, since by definition 
$[\hat{x},\hat{p}] \neq [\hat{X},\hat{P}]$, then the transformation $\hat{X}(\hat{x})$, $\hat{P}(\hat{p})$ is non-canonical, namely the correspondent Poisson brackets are not preserved. Therefore, the GUP is really describing new physics, doesn't simply describe the usual physics with different variables. A particularly useful example of such non-linear representations is the following. Consider (with $\beta_0 = \beta/m_p^2 c^2$)
\be
\left\{\begin{array}{l} 
\hat{X} = \hat{x} \\ \hat{P} = \hat{P}(\hat{p}) 
\end{array}
\right. \,; \quad {\rm then \ from} \quad
\left\{\begin{array}{l} 
[\hat{X}, \hat{P}] = i\hbar (1+ \beta_0 \,\hat{P}^2) \\ {[\hat{x}, \hat{p}] = i\hbar}
\end{array}
\right. \,, \quad {\rm follows} \quad
\hat{P} = \frac{1}{\sqrt{\beta_0}} \tan (\sqrt{\beta_0} \,\hat{p})\,.
\ee 
If we use the standard position representation of the low energy momentum $\hat{p} = -i\hbar\partial_x$ , then the Hamiltonian written with the high energy variables is actually equivalent to a low energy canonical Hamiltonian containing an infinite derivative kinetic term
\be
\hat{H} = \frac{\hat{P}^2}{2m} + U(\hat{X}) = \frac{1}{2m\beta_0} [\tan(-i\hbar\sqrt{\beta_0}\partial_x)]^2
+ U(\hat{x})\,.
\label{H}
\ee
Of course, the corrections at the first order in $\beta$ to the standard quantum mechanical results are computed in perturbation theory by expanding the tangent function in \eqref{H}. So the Hamiltonian governing a GUP-deformed quantum system will be (at the first order in $\beta$)
\be
\hat{H} = \frac{\hat{P}^2}{2m} + U(\hat{X}) = \frac{\hat{p}^2}{2m} + U(\hat{x}) + \beta_0 \frac{\hat{p}^4}{3m}\,.
\ee	
Using the above Hamiltonian, various different systems have been re-computed, and, by comparison with the experimental data, the following upper bounds for the deformation parameter $\beta$ have been obtained:
\begin{center}
\begin{table}[ht]
\caption{Upper bounds on $\beta$.}
\begin{tabular}{c|c}
  \hline\hline
    Physical framework   & $\beta <$ \\ \hline             
    ${}^{87}$Rb cold-atom-recoil experiment  & $10^{39}$ \\ \hline
			Charmonium levels    & $10^{34}$ \\
             Energy difference in Hydrogen levels $1S-2S$  & \\ \hline
						Electroweak measurement & $10^{34}$ \\ \hline
						Lamb shift    &  $10^{20}$ \\ \hline
						Micro and nano mechanical   & $10^{12}$ \\
              oscillators (masses $\sim \mpl$)  & \\ \hline
      Sapphire mechanical resonator    & $10^{6}$ \\ \hline
 \end{tabular}
\label{Tab1}
\end{table}
\end{center}
Such large upper bounds should not impress the reader. Somehow, we were to expect them, given the extremely large suppression of the deforming term in $\beta$, due to the smallness of the coefficient 
$1/m_p^2c^2=4\ell_p^2/\hbar^2$. As we said in the previous Section, an exact theoretical calculation of $\beta$ has been performed in Ref.~\cite{SLV}, and it was found $\beta=82\pi/5$,
namely $\beta$ is of order $10$ (as already suspected on different theoretical grounds). A simple glimpse to the experimental bounds reported in Table I immediately testifies of the large (or perhaps huge) gap that experimental investigations still have to cover in order to arrive to test the theoretical value.

\section{Hawking temperature, GUP, and black hole remnants}

Among the many things that can be recomputed through the GUP there is the Hawking temperature of an evaporating black hole. Let us first derive the standard Hawking formula directly from the usual Heisenberg principle~\cite{Scardigli95}. The argument proceeds as follows: suppose you have a bunch of Hawking photons just outside the event horizon of a Schwarzschild black hole of mass $M$. Further, suppose the photons are in a semiclassical regime, namely their wavelength is large, so that the equipartition theorem holds, and the average energy of a Hawking photon is given by $E = k_B T$. If the photons are close to the event horizon, then the uncertainty in their position is of the order of the geometrical size of the horizon, namely $\delta x \simeq 2\mu R_g$, where $\mu$ is a parameter that will be adjusted in order to match the actual Hawking temperature. Then, since relation 
\eqref{deltaxH} between $\delta x$ and $E$ should hold, we can also write 
\be
2\mu R_g = \frac{hc}{k_B T}
\ee
or
\be
T = \frac{hc}{2\mu k_B R_g} = \frac{hc^3}{4\mu k_B GM} = \frac{\hbar c^3}{8\pi k_B GM}\,,
\ee
where the last equality holds if we choose $\mu=4\pi^2$,  and of course $\hbar=h/2\pi$. The above derivation uses the standard Heisenberg relation \eqref{deltaxH}. But we can now repeat exactly the above steps by using instead the GUP relation \eqref{deltaxE}. Then one ends up with a deformed Hawking temperature $T$ satisfying the equation
\be
M \ = \ \frac{\hbar c^3}{8\pi k_B GT} \ + \ \beta\,\frac{k_B T}{8\pi^2 c^2}\,.
\label{MT}
\ee
Relation \eqref{MT} is directly derived from the GUP, and of course reduces to standard Hawking formula for 
$\beta \to 0$ . Relation \eqref{MT} can be easily inverted, and then expanded to the first order in $\beta$, so obtaining
\be
T \ = \ \frac{\hbar c^3}{8\pi k_B GM} \left(1 \ +  \ \frac{\beta}{4\pi^2} \frac{m_p^2}{M^2} \ + \ \dots\right)\,.
\ee
The physical predictions of relation \eqref{MT} are striking in many different sectors. 
Astrophysics and cosmology, among others, are fields deeply affected by the consequences of Eq.\eqref{MT}. Consider for example the diagram in Fig.2.
\begin{figure}[t]
	\includegraphics[scale=1.2]{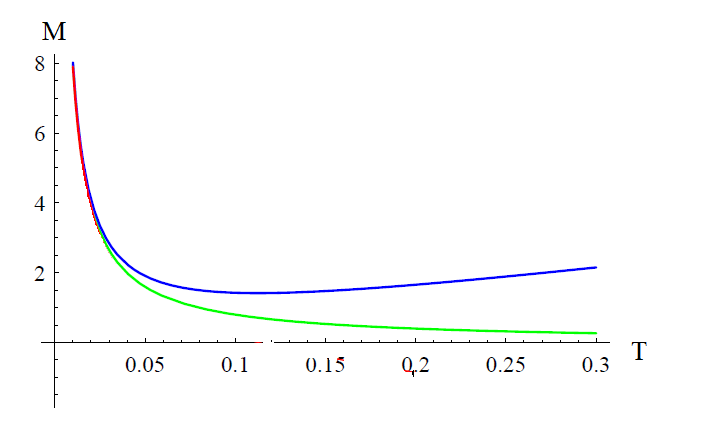}
	\centering
	\protect\caption{Mass versus Temperature (in Planck units) for the quantum (micro) black hole.}
	\label{fig2}
\end{figure}
The green (bottom) line represents the standard mass/temperature relation of Hawking for a Schwarzschild black hole, while the blue (top) line is the GUP-deformed mass/temperature relation, Eq.\eqref{MT}. As we see, the usual Hawking formula predicts that the black hole evaporation goes on until a final \textit{zero} mass and an \textit{infinite} final temperature. On the contrary, according to the GUP-deformed formula, Eq.\eqref{MT}, the mass/temperature diagram has a minimum. So the evaporation starts with a large mass $M$, and halts when $M$ reaches a minimum mass 
$M_{min} \sim m_p$ and a correspondent maximum \textit{finite} temperature $T_{max} \sim T_p$ (being $T_p$ the Planck temperature, defined as $k_B T_p = E_p$). The use of the GUP has the great advantage of avoiding an unphysical infinite final temperature. Actually, a more refined analysis reveals that the temperature of the (micro) black hole drops to zero in the very last phases of the evaporation. So, GUP predicts the formation of small inert \textit{remnants} as final product of the black holes evaporation (see Ref.~\cite{ACS}). These remnants do not have any thermodynamic interaction with the environment. This is confirmed by their specific heat, which drops to zero at the end of the evaporation, when  $M \to M_{min}$. Remnants interact with the environment just only because they gravitate, through their final \textit{strictly positive} mass.

\section{Black hole remnants and dark matter}

Exactly for the above specific property, black hole remnants have been considered, since the moment of their introduction, as very good candidates for being a major constituent of dark matter. 
As it is well known by many decades~\cite{Rubin80}, there is an anomaly in the rotational curves of almost all the galaxies. By measuring the orbital parameters (speed, in particular) of individual stars in a given galaxy, astronomers are able to reconstruct the amount of gravitating mass necessary to generate that particular orbit. They systematically find that the gravitating mass able to reproduce the observed stellar orbits is larger (by far) than the total mass of luminous matter contained in the galaxy. So, admitting the validity of Newtonian/Einsteinian gravity, there should be present a large quantity of mass, not visible, not interacting electromagnetically, which however gravitates, and it is therefore able to explain the observed data. This invisible matter is called \textit{dark matter}. Incidentally, the other possible explanation of the dilemma, for those who don't like the dark matter hypothesis, is to invoke a modification of the Newtonian/Einsteinian gravity on galactic scales~\cite{Milgrom,Bekenstein}. 
\begin{figure}[t]
	\includegraphics[scale=0.2]{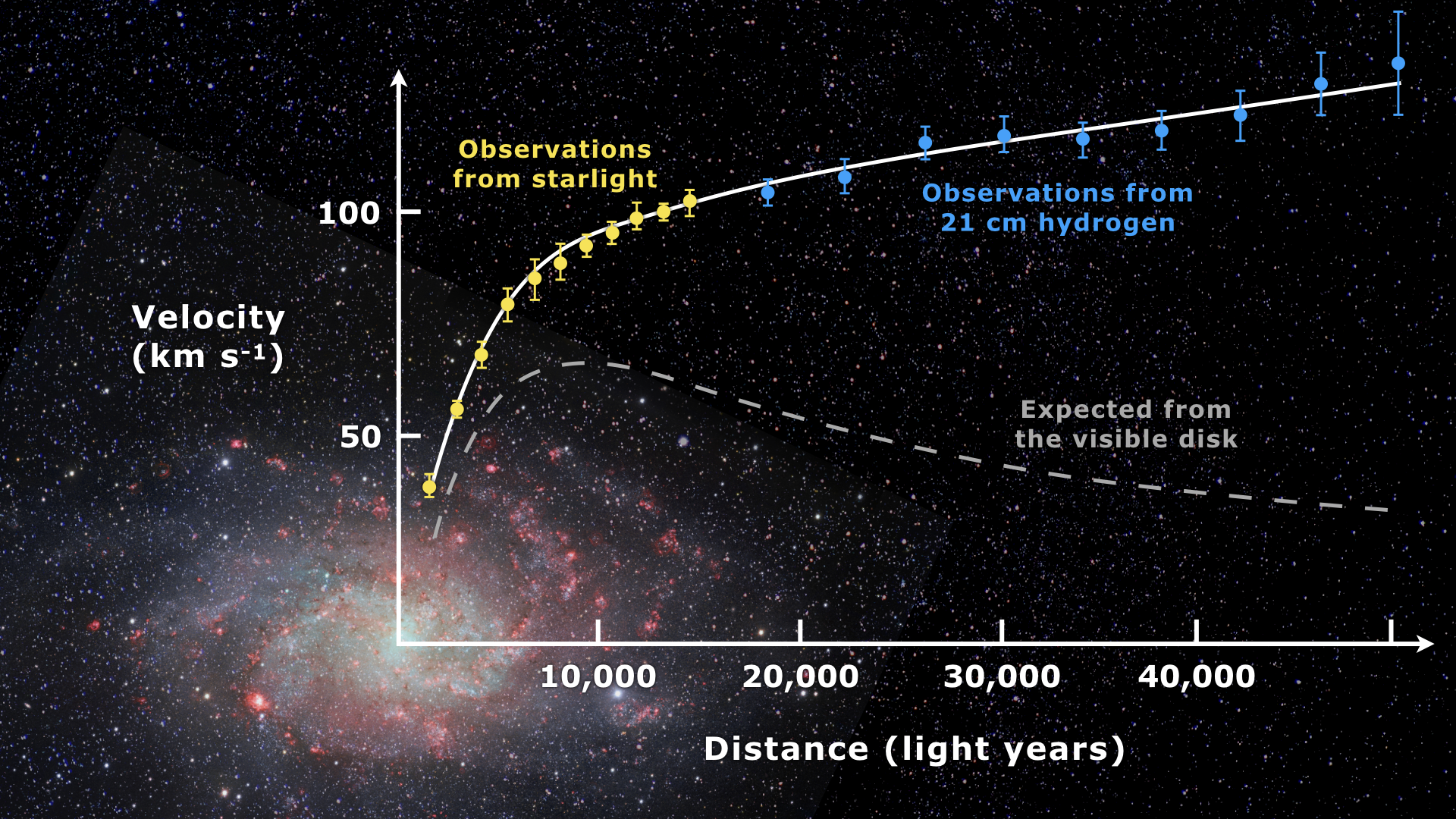}
	\centering
	\protect\caption{Anomaly in the rotational curves of galaxies.}
	\label{fig3}
\end{figure}

During the years, many different hypothesis have been proposed for the specific constituents of the dark matter. For example, several authors pointed out that if, as it seems, neutrinos have a mass (small, but not zero), then perhaps 'clouds' of neutrinos, gravitationally bound to the galactic disks, may constitute the invisible 'dark' mass responsible for the anomalous rotational curves. Other authors propose the existence of some exotic form of matter, as axions, or maybe supersymmetric partners of ordinary particles, in order to explain dark matter. 
It is therefore now clear why black hole remnants, with their specific property of having a non-zero mass, and no other interacting features, have been considered from the beginning as very good natural candidates as dark matter constituents~\cite{ACS}.

\section{Conclusions}

This work presents a general, pedagogical introduction to the idea of how and why a fundamental principle like the Heisenberg uncertainty should be modified when the gravitational interaction is taken into account. Hence, we discuss the so called Generalized Uncertainty Principle (GUP), and, among the many different ways, we introduce it through the idea of possible formation of micro black holes during a measurement (i.e. scattering) process. Only few of the many striking consequences of the GUP have been mentioned (also because, many of them are still a matter of debated investigations). In particular we showed how the GUP affects a generic Hamiltonian, which in turn serves to compute modifications of several basic quantum phenomena. By comparing theoretical results with experimental numbers, upper bounds on the parameter $\beta$ governing such modifications can be found. Finally, we discussed the idea of black hole remnants, one of the most distinctive predictions of the GUP applied to the Hawking evaporation of quantum black holes. Consequences of this idea for the dark matter problem are briefly mentioned. In this way, we have given a general panorama of the GUP motivations, and of some of its physical consequences.


%
%
%
\end{document}